\documentclass{iopart}  
\usepackage{epsfig,subfigure}

\newcommand{\be}{\begin{equation}}
\newcommand{\bel}[1]{\begin{equation}\label{#1}}
\newcommand{\ee}{\end{equation}}
\newcommand{\bea}{\begin{eqnarray}}
\newcommand{\ba}{\begin{array}}
\newcommand{\eea}{\end{eqnarray}}
\newcommand{\ea}{\end{array}}

\begin{document}
\jl{1}

\comment{Comment on `Garden of Eden states in traffic model revisited'}
\author{Andreas Schadschneider\dag\ and Michael Schreckenberg\ddag}

\address{\dag\ Institut f\"ur Theoretische Physik, Universit\"at zu 
K\"oln, D-50937 K\"oln, Germany}
\address{\ddag Theoretische Physik FB 10, Gerhard-Mercator-Universit\"at
Duisburg, D-47048 Duisburg, Germany}

\pacs{05.70.Ln, 45.70.Vn, 05.40.-a, 02.50.Ga}
\submitted
\begin{abstract}
Recently, Huang and Lin suggested a combination of two successfull
mean-field theories, the 2-cluster approximation and paradisical mean-field, 
for the Nagel-Schreckenberg cellular automaton model of traffic flow.
We argue that this new approximation is inconsistent since it
violates the Kolmogorov conditions.
\end{abstract}


In a recent work \cite{Huang} Huang and Lin have studied a 
combination of two mean-field theories, the 2-cluster 
approximation \cite{cluster1,SSNI,AS} 
and paradisical mean-field (PMF) \cite{GoE,AS}, 
for the Nagel-Schreckenberg cellular automaton model of traffic 
flow \cite{NaSch} (for a review, see \cite{CSS99}). 
The suggested combined theory yields results for the flow-density relation 
which are worse compared to Monte Carlo simulations than those of each 
individual approximation alone. The authors concluded that the success
of paradisical mean-field theory is accidental and cannot be improved
systematically, in contrast to the cluster approximation.
In this comment we will show that the combined theory as suggested
by Huang and Lin is inconsistent since it violates the elementary
Kolmogorov conditions. This also invalidates the conclusions about
the success of paradisical mean-field theory.

In the 2-cluster approximation the probability $P(\tau_1,\ldots,\tau_L)$
to find the system in the state $(\tau_1,\ldots,\tau_L)$ is
factorized into 2-site probabilities $P_{\tau,\tau'}$:
\begin{equation}
P(\tau_1,\ldots,\tau_L)\propto P_{\tau_1,\tau_2}P_{\tau_2,\tau_3}\cdots
P_{\tau_{L-1},\tau_L}P_{\tau_L,\tau_1}.
\end{equation}
Here we assume translational invariance so that the probabilities 
$P_{\tau,\tau'}$ do not depend on the position.

An important consequence are the so-called Kolmogorov consistency
conditions \cite{kolmogorov} which for the 2-cluster approximation
read
\begin{equation}
\sum_{\tau_2}P_{\tau_1,\tau_2} = P_{\tau_1}=\sum_{\tau_2}P_{\tau_2,\tau_1}
\label{eq_kolmo}
\end{equation}
where $P_{\tau}$ is the probability to find a cell in state $\tau$.
Especially for $\tau=x$, denoting an empty cell in the notation of
\cite{Huang}, we have $P_x=1-\rho$ where $\rho$ is the total density
of cars. Since also $P_1+P_2=\rho$, instead of eq.~(2) in \cite{Huang}
one has more precisely
\begin{eqnarray}
\rho &=& P_{1x}+P_{11}+P_{12}+P_{2x}+P_{21}+P_{22}\label{eq_rho1}\\
&=& P_{x1}+P_{11}+P_{21}+P_{x2}+P_{12}+P_{22}
\end{eqnarray}
and
\begin{eqnarray}
1-\rho &=& P_{xx}+P_{x1}+P_{x2}\\
&=& P_{xx}+P_{1x}+P_{2x}.\label{eq_rho4}
\end{eqnarray}


We have analyzed numerically the equations given in the Appendix
of \cite{Huang} and found that these consistency conditions are
violated. In fact we 
found that the results depend strongly on the iteration scheme used. 
If one adjusts the normalization ${\cal N}$ during the iteration to ensure 
the normalization of probabilities (eq.~(1) of \cite{Huang}) then
the numerical solution converges to $\rho=0$, even for initial values
with $\rho\neq 0$.
On the other hand, if one uses eq.~(1) of \cite{Huang} to eliminate one
of the equations in the Appendix, the normalization condition is always
satisfied during iteration. One then can use ${\cal N}$ to fix the density
$\rho$ according to eq.~(2) of \cite{Huang}. In this case we find that 
the Kolmogorov consistency conditions (\ref{eq_rho1})-(\ref{eq_rho4})
are violated and in general
$\sum_{\tau}(P_{1\tau}+P_{2\tau})\neq \rho\neq \sum_{\tau}
(P_{\tau 1}+P_{\tau 2})$.
It even seems that no solution exists, at least in certain density regimes.

In order to exclude the possibility that these problems are related to a
mistake or typo in the set of equations given in \cite{Huang} we have done
an independent calculation. However, we find it much more convenient
not to use the update ordering R2-R3-R4-R1 that was used in 
\cite{cluster1,SSNI} and adopted in \cite{Huang}. Instead 
the original update order R1-R2-R3-R4 \cite{unpublished} is preferable
since it allows a much easier identification of the GoE states and avoids
the introduction of the weighting ${\cal W}$. It also turns out
that this approach is numerically much more stable since no terms 
proportional to ${\cal N}^2$ (where ${\cal N}$ is the normalization) appear.
For our system of equations we find the same type of behaviour as described 
above for the equations of \cite{Huang}.

The reason for this failure is that the modified equations after
elimination of the GoE states do not automatically guarantee the
conservation of cars, in contrast to the original 2-cluster equations.
It is not possible to satisfy the normalization condition
$\sum_{\tau_1,\tau_2}P_{\tau_1,\tau_2} = 1$ 
and the Kolmogorov conditions (\ref{eq_kolmo}) by just introducing
one normalization factor ${\cal N}$.
One could try to use normalizitions ${\cal N}_{\tau,\tau'}$ depending on
the state, but this would make the system of equations even
more complicated than the 3-cluster approximation \cite{cluster1,SSNI,AS} 
which already treats most GoE states correctly.

Finally we want to comment on the conclusions of \cite{Huang}. 
We believe that the combined approach -- if successful -- must be
considered as an improvement of the 2-cluster approach, not of PMF
as suggested in \cite{Huang}. If it would be possible to carry out
such an approximation consistently it would most certainly lead to an 
improvement of the 2-cluster results. 
In fact the main point of the PMF is not the quantitative agreement
with simulations but the fact that it allows to determine the origin
of correlations and thus to improve the understanding of the underlying
physics. This was already emphasized in \cite{GoE,AS}.





\section*{References}
\begin{thebibliography}{99}
\bibitem{Huang}
Huang D and Lin Y 2000 J.\ Phys.\ A~{\bf 33}, L471 

\bibitem{cluster1}
Schadschneider A and Schreckenberg M 1993 J.\ Phys.\ A~{\bf 26}, L679

\bibitem{SSNI}
Schreckenberg M, Schadschneider A, Nagel K and Ito N 1995 
Phys.\ Rev.\ E~{\bf 51}, 2939

\bibitem{AS}
Schadschneider A 1999 Eur.\ Phys.\ J.\ B~{\bf 10}, 573 

\bibitem{GoE}
Schadschneider A and Schreckenberg M 1998 J.\ Phys.\ A~{\bf 31}, L225 

\bibitem{NaSch}
Nagel K and Schreckenberg M 1992 J. Phys. I France {\bf 2}, 2221

\bibitem{CSS99}
Chowdhury D, Santen L and Schadschneider A 2000
Phys.\ Rep.\ {\bf 329}, 199 

\bibitem{kolmogorov} Gutowitz H A, Victor J D and Knight B W 1987
Physica {\bf 28D}, 18 

\bibitem{unpublished}
Schadschneider A, unpublished results

\end{thebibliography}
\end{document}